\documentclass[10pt,article]{IEEEtran}

\usepackage{amsmath}
\usepackage{amsfonts}
\usepackage{graphicx}
\usepackage{float}
\usepackage[square,numbers]{natbib}
\usepackage{dirtytalk}
\usepackage{multirow}
\usepackage{caption} 
\author{Michael Filletti, Aaron Grech}
\title{Using News Articles and Financial Data to predict the likelihood of bankruptcy}

\begin{document}

\maketitle

\section{Introduction} \label{sec:intro}
Over the past decade, millions of companies have filed for bankruptcy. This has been caused by a plethora of reasons, namely, high interest rates, heavy debts and government regulations. When a company goes bankrupt, one might instantly think that this will affect the owners and workers of the company. However, the effect spreads far wider than that. Clients, suppliers, shareholders, auditors and any related companies are all influenced by this event. A company with various related companies with a strong foothold in a market can even influence the economy of an entire country \citep{charitou2004}. With markets becoming increasingly competitive, and more advanced statistical techniques and data being available to us, the ability to evaluate the financial health of an organization has become a more and more interesting issue to us \citep{balcaen2006}. Although financial ratios have been shown to produce strong results \citep{altman1968}, it is clear that quantitative variables are not the only factor influencing companies going bankrupt. Factors such as economic climate, new technologies and political climate all have a big influence on the likelihood of a company closing down. Quantitative variables such as accounting ratios do not take this into consideration.
\newline
One of the aims of this assignment is to provide a framework for company bankruptcy to be predicted by making use of financial figures, provided by our external dataset, in conjunction with the sentiment of news articles about certain sectors. News articles are a cheap and widely available source of data, providing information almost at real time. This makes them a very powerful source of information to be used, to show the growth, or decline, of an industry. Another advantage of news articles is that they give an overall view of the situation with the industry, both at a micro and macro level. The second aim of this assignment is to evaluate the performance of such an approach and find areas of improvement, to work our way towards making this a widely available solution to a problem that has existed for decades. Due to the fixed time frame that is being worked in, and the amount of work required, the results are not expected to be groundbreaking. Results are also not expected to be of a high quality due to the provided financial dataset being pseudonymised, which means that there is an inherent challenge to match the sectors for which news articles will be obtained and the financial data sectors. For these reasons, the expectation is to build something that gives promising results, and more importantly, has plenty of potential to be used to predict bankruptcy in a smarter way. The methods and sources of information are almost all easily and freely available, meaning that this work can be easily replicated and improved. That said, this assignment should give a good idea of the quality of sentiment that is obtained from newspaper articles, as well as identifying the challenges and potential areas of improvement.
\section{Related Research} \label{sec:research}
The problem of predicting bankruptcy has been attempted to be solved for many decades, the first real breakthrough came from \citet{altman1968}, who suggested a multivariate discriminant analysis, building on the univariate discriminant analysis that \citet{beaver1966} proposed a few years earlier. Altman suggested the use of a Z-Score, calculated using five financial ratios, to predict bankruptcy. The calculation of the Z-Score is shown below:
\begin{equation}
    Z=1.2\cdot A + 1.4\cdot B + 3.3\cdot C + 0.6\cdot D + 1.0\cdot E
\end{equation}
The five variables in the equation represent the following:
\begin{itemize}
    \item $A={\mathrm{working\;capital}}\;/\;{\mathrm{total\;assets}}$
    \item $B={\mathrm{retained\;earnings}}\;/\;{\mathrm{total\;assets}}$
    \item $C={\mathrm{earnings\;before\;interest\;and\;tax}}\;/\;{\mathrm{total\;assets}}$
    \item $D={\mathrm{market\;value\;of\;equity}}\;/\;{\mathrm{total\;liabilities}}$
    \item $E={\mathrm{sales}}\;/\;{\mathrm{total\;assets}}$
\end{itemize}
\bigskip
If $Z>3$, a company is unlikely to go bankrupt, while in the case that $Z<1.8$ the company is likely to go bankrupt. For values of $Z$ where $1.8<Z<3$, this is referred to as the grey area. These ratios have been used in vast amounts of research, whilst Altman himself has also provided improved variations to the model over the years \citep{altman2014, altman2000, altman1977}. Since then, various algorithms have been used to enhance the method of prediction. Neural networks have often been used to attempt to solve this problem. \citet{odom1990} were one of the first researchers to propose making use of a neural network to tackle this problem. This literature compares the neural network against a more traditional multivariate discriminant analysis, with promising results. A well known book presenting the use of this method was written by \citet{trippiturban1992}. Although neural networks are a popular solution, other methods have been applied to this problem. One method suitable for predicting bankruptcy is survival analysis, investigated by \citet{parker2002} and \citet{gepp2008}. The technique is also attempted by \citet{lee2014}, who makes use of the Cox Proportional Hazard Model to evaluate the technique. Only a few financial ratio variables are used to construct the model, and fairly strong results are achieved, with an overall accuracy of classification of 87.93\%. \citet{lin2001} investigate the use of various techniques to predict bankruptcy, finding that Decision Trees also provide good results. \citet{kumar2007} provide a general overview of the work done from 1968 up to 2005. Various techniques that tackled this problem are reviewed, namely the aforementioned neural networks and decision trees.
\\
\newline
One of the biggest challenges in this experiment will be to actually quantify the sentiment on a particular sector from the news articles. \citet{liu2012} defines Sentiment Analysis as \say{the field of study that analyzes people?s opinions, sentiments, evaluations, appraisals, attitudes, and emotions towards entities such as products, services, organizations, individuals, issues, events, topics, and their attributes}. With the rise of social media and the ever-increasing amount of information available online, sentiment analysis has become more and more relevant. It is inherently, a very difficult task, due to the wall that exists between the writer and the reader, whereby different people have different means of expressing themselves. This task can be undertaken in two different ways, one of which is the machine learning-based approach, the other being the lexicon-based approach \citep{li2010,chaovalit2005}. In the machine learning-based approach, a number of labelled phrases are used, classified as sentiments. These are used as the training set, while any new phrases or sentences are classified based on the labelled training set. On the other hand, the lexicon-based approach makes use of a dictionary that contains words classified as certain emotions, and documents are scored based on that dictionary. \citet{chaovalit2005} compare both methods to classify movie reviews, specifying that it was a challenge for both approaches. \citet{loughran2011} propose a dictionary to be used for financial texts, considering four different types of emotions: Positive, Negative, Uncertainty and Litiguous. \citet{li2014} uses this dictionary to investigate the relationship between stock prices and financial news articles. In the article, it is concluded that models using sentiment analysis outperform those using a bag-of-words model. Due to the inherent link between stock prices and financial news, it is quite common for researchers to investigate this. \citet{hajek2015} investigate the correlation between the sentiment of financial reports and stock returns, once again making use of the Loughran-McDonald dictionary, obtaining positive results.
\\
\newline
Another major challenge in this experiment is dealing with the class imbalance problem. Class imbalance occurs when we have a classification problem, in which one class is significantly more common than another. In this case, the number of bankrupt companies are significantly less rare than those that are still running. In our dataset, we tend to have a 95-5 split with some variation. \citet{le2018} discuss this in the context of the bankruptcy problem, proposing a cluster based boosting algorithm to tackle this problem. \citet{zikeba2016} explain that bagging and boosting were the first attempts at solving this class imbalance problem. In their experiment, they make use of synthetic feature generation to oversample the bankrupt companies. An alternative to this is the Synthetic Minority Over-sampling Technique (SMOTE), proposed by \citet{chawla2002}. This is used by \citet{sun2018} and \citet{kim2015} to tackle the bankruptcy problem. \citet{fan1999} adapted boosting to create AdaCost, a misclassification cost-sensitive boosting method. These class imbalance techniques have been consistently improved, namely by the creation of the Extreme Gradient Boosting technique \citep{chen2015}, and by ensemble classifiers. These methods have achieved strong results in the very domain we are interested in \citep{zikeba2016,nanni2009}.
\\
\newline
In their research, \citet{jo2016} propose a bankruptcy prediction model for Korean construction firms, making use of both qualitative and quantitative data. A domain-specific sentiment lexicon is used to extract the sentiment from the economic news articles. After obtaining all the news articles and carrying out various preprocessing techniques, 367 terms were obtained. Each of these terms were assigned a specific category depending on their score. The score is calculated by considering the number of positive news articles a word is in and the number of negative news articles a word is in, as well as the Construction Business Survey Index (CBSI). Multiple models were tested for each time period, whilst also taking into consideration various different sentiment variables, namely the average sentiment score or the difference between the percentage of positive articles and negative articles. The machine learning technique used is the artificial neural network. The results from this experiment were found to be positive, with clear indications that the qualitative variables combined with the quantitative variables aided in predicting bankruptcy. In particular, it was found that negative sentiment was able to predict bankruptcy better than other sentiments. The clear limitation with this exercise was the very limited and specific lexicon used. In this exercise, the lexicon was generated manually, by making use of the CBSI. A business index, however, is not always available, making this a very specific solution to the issue. For this problem to be tackled at a scalable level, one would need a global method of tackling this sentiment analysis problem.

\section{Solution Design and Implementation} \label{sec:design}
\citet{jo2016} and \citet{hajek2015} propose interesting solutions to the problem, that we will be building upon. As mentioned in section \ref{sec:research}, although their results were quite positive, the inherent problem with their solution was making use of a domain-specific lexicon and index to carry out the key part of the exercise, which are not always available. In this exercise, we aim to evaluate the performance of a solution that can be applied to all domains, while offering good performance.
\newline
For this experiment, two datasets were used, one containing the financial data of companies, and the other being the lexicon containing the sentiment category of various words. The financial dataset was provided by the Malta Information Technology Agency (MITA), and is split in two, one dataset containing data relevant to the balance sheet, and another containing data relevant to income statements. The balance sheets contain 253 fields, while the income sheets are made of 257 fields, both sheets containing relevant financial data. In addition to this, there are 10 fields containing general company information such as the NACE code, the year of assessment (2004-2017), status (whether still running or not) and so on. The financial datasets follow the Tax Index of Financial Data (TIFD) format. This is a format that was proposed to begin to regulate reporting requirements for tax purposes. Such regulations should be taken advantage of to help solve problems such as the one we are attempting.
\\
\newline
As one can tell, the financial dataset is enormous, containing a huge amount of features and instances. For this reason, missing values are to be expected. Rather than delving into this issue and attempting to find the ideal combination of features through feature selection, or applying some other dimensionality reduction technique, the variables related to the Z-score of \citet{altman1968} will be used. This has been proved to provide positive results, as discussed in section \ref{sec:research}. All the variables required to calculate this Z-Score are provided within our dataset. The only issue comes with the $D$ variable, which represents ${\mathrm{market\;value\;of\;equity}}\;/\;{\mathrm{total\;liabilities}}$. This variable only exists for public companies. For private companies, this is equivalent to 0. However, there is an alternative, revised Z-score, that makes use of the ${\mathrm{book\;value\;of\;equity}}\;/\;{\mathrm{total\;liabilities}}$ instead \citep{altman2000}. The formula for this alternative Z-Score $Z'$ is shown below:
\begin{equation}
    Z'=0.717\cdot A + 0.847\cdot B + 3.107\cdot C + 0.420\cdot D + 0.998\cdot E
\end{equation}
The five variables in the equation represent the following:\newline 
\begin{itemize}
    \item $A={\mathrm{working\;capital}}\;/\;{\mathrm{total\;assets}}$
    \item $B={\mathrm{retained\;earnings}}\;/\;{\mathrm{total\;assets}}$
    \item $C={\mathrm{earnings\;before\;interest\;and\;tax}}\;/\;{\mathrm{total\;assets}}$
    \item $D={\mathrm{book\;value\;of\;equity}}\;/\;{\mathrm{total\;liabilities}}$
    \item $E={\mathrm{sales}}\;/\;{\mathrm{total\;assets}}$
\end{itemize}
\bigskip
If $Z'<1.21$ then the company is said to have a high likelihood of bankruptcy, while vice-versa if $Z'>2.90$. We utilize both scores in our experiments, using the revised model for those companies that are not publicly traded, and the original model for companies that are. Any instances with financial ratios that returned infinite values or could not be calculated were excluded from our final dataset.
\newline
To carry out the sentiment analysis, the lexicon-based approach is taken. In this scenario, a dictionary of labelled, or scored words is necessary. \citet{loughran2011} provide a dictionary specifically to be used in the financial context, building upon what was done by the Harvard General Inquirer\footnote{http://www.wjh.harvard.edu/~inquirer/homecat.htm}. As was mentioned in section \ref{sec:research}, this has been used in multiple research problems, mainly used on financial news articles or annual reports. We attempt to use this on general news articles, which will be somewhat different in terminology in comparison to specific financial news articles. In this assignment, a newer, updated version of the Loughran and McDonald dictionary (2016) is used.
\newline
The textual data is scraped from the Maltese news portal Times of Malta\footnote{http://www.timesofmalta.com}, with data being scraped per year for the industry sectors considered. The industry sectors that will be considered are:
\begin{itemize}
    \item iGaming
    \item Pharmaceuticals
    \item Aviation
    \item Tourism
\end{itemize}
The sectors were used as keywords to scrape and obtain related information from the site on a year by year basis. This means that we had a collection of articles per year per sector. The years of articles scraped were from 2013 to 2016. Once the text is made available, preprocessing must occur for us to be able to extract some value from it. This involves carrying out techniques such as tokenization and lemmatization. Any text that is common in all articles and represents no meaningful value is excluded from the final collection of terms. Tokenization is performed, which segments the text into words, followed by converting these words to the lower case, converting numbers to words, removing punctuation and removing stopwords. Stopwords refer to words that are very common to the language and offer no sentiment value at all, such as a, the, there, and so on. Finally, lemmatization is used to remove any issues with the tense of the word. 

All the techniques mentioned above were carried out in \texttt{Python 3.6} using the Natural Language Toolkit ($\texttt{NLTK}$). These same techniques were carried out for the Loughran and McDonald dictionary, so that we will have a like for like comparison when evaluating our news articles.
\\
\newline
Below is a table illustrating the results of using the Loughran and McDonald dictionary on our news articles.
\begin{table}[H]
\centering
\begin{tabular}{ |c|c||c|c|c| }
    \hline
        Sector & Year & Positive\% & Negative\% & PositiveToNegative \\ \hline
    \hline
        \multirow{4}{*}{Aviation}& 2013 & 0.92\% & 1.41\% & 0.653 \\ 
        & 2014 & 0.83\% & 1.60\% & 0.518 \\ 
        & 2015 & 0.82\% & 1.50\% & 0.548 \\ 
        & 2016 & 0.97\% & 1.43\% & 0.676 \\ \hline
        \multirow{4}{*}{iGaming}& 2013 & 1.19\% & 1.50\% & 0.790 \\ 
        & 2014 & 1.21\% & 1.48\% & 0.814 \\ 
        & 2015 & 1.21\% & 1.33\% & 0.911 \\ 
        & 2016 & 1.17\% & 1.50\% & 0.783 \\ \hline
        \multirow{4}{*}{Pharmaceuticals}& 2013 & 1.15\% & 1.35\% & 0.848 \\ 
        & 2014 & 1.12\% & 1.54\% & 0.730 \\ 
        & 2015 & 1.11\% & 1.26\% & 0.882 \\ 
        & 2016 & 1.12\% & 1.45\% & 0.773 \\ \hline
        \multirow{4}{*}{Tourism}& 2013 & 1.14\% & 1.25\% & 0.910 \\ 
        & 2014 & 1.12\% & 1.28\% & 0.879 \\ 
        & 2015 & 1.11\% & 1.30\% & 0.854 \\ 
        & 2016 & 1.03\% & 1.34\% & 0.765 \\
     \hline
\end{tabular}\par
\caption{Sentimental Variables found by our dictionary and to be used to predict bankruptcy}
\label{Tab:sent}
\end{table}
As can be seen from the table, each sector has similar levels of rates. The iGaming, Pharmaceutical and Tourism industries tend to have higher levels of positive word rates, while the Aviation industry had noticeably lower rates than the other three sectors. The Aviation industry also had a low percentage of Positive words and a high percentage of Negative words. It is interesting to note that articles related to the iGaming industry had a relatively high percentage of both Positive and Negative words. This could perhaps be related to the fact that it is a volatile and controversial industry. The Tourism industry suffered a decline in terms of Positive to Negative words, with the number of Positive\% words decreasing and Negative\% words increasing. The number of tourists over these years has actually been increasing, as per the Malta Tourism Authority (MTA), however it would be interesting to look into the revenues made by relevant companies within the touristic sector. In addition, we should also note that these sentiments can take some time to be seen in hard figures, so a drop in sentiment may be a warning for future years. Another possibility is that it may signal the need for the articles scraped to be filtered. The Pharmaceuticals industry had a fairly stable level of sentiment, with the sentiment scored altering from year to year. In general, this the most stable and one of the brightest sectors in Malta \citep{maltaenterprise}, so such scores tend to make sense.
\newline
One issue that should be mentioned within our experiment is the amount of irrelevant articles that may exist within the searched keyword. This could have an impact on our results, however, due to time constraints, could not be addressed. As previously mentioned, a potential solution is to filter out articles that are not relevant. This could be an issue due to our source of data, which is a general news portal, not specialised in business articles. A second potential issue within this experiment is the inherent nature of news articles being unbiased, meaning that the levels of positive and negative words across sectors are fairly similar. That said, despite such issues it is interesting to note the different variations we see across various sectors, and the scores do tend to make sense, as discussed previously. 
\\
\newline
The variables selected to be used in this experiment are:
\begin{itemize}
    \item $\mathbf{A}={\mathrm{working\;capital}}\;/\;{\mathrm{total\;assets}}$
    \item $\mathbf{B}={\mathrm{retained\;earnings}}\;/\;{\mathrm{total\;assets}}$
    \item $\mathbf{C}={\mathrm{earnings\;before\;interest\;and\;tax}}\;/\;{\mathrm{total\;assets}}$
    \item $\mathbf{D}={\mathrm{market\;value\;of\;equity}}\;/\;{\mathrm{total\;liabilities}}$
    \item $\mathbf{D'}={\mathrm{book\;value\;of\;equity}}\;/\;{\mathrm{total\;liabilities}}$
    \item $\mathbf{E}={\mathrm{sales}}\;/\;{\mathrm{total\;assets}}$
    \item $\mathbf{Z}=1.2\cdot A + 1.4\cdot B + 3.3\cdot C + 0.6\cdot D + 1.0\cdot E$
    \item $\mathbf{Z'}=0.717\cdot A + 0.847\cdot B + 3.107\cdot C + 0.420\cdot D' + 0.998\cdot E$
    \item \textbf{Negative\%}: Percentage of Negative words (defined by \citet{loughran2011} in all news articles)
    \item \textbf{Positive\%}: Percentage of Positive words (defined by \citet{loughran2011} in all news articles)
    \item \textbf{PositiveToNegative}: Ratio of Positive words to Negative words (defined by \citet{loughran2011} in all news articles)
    \item \textbf{Status}: Indicates whether a company is bankrupt or not
\end{itemize}
These variables were chosen based on the work of \citet{altman1968,altman2000}, as well as the work of \citet{jo2016}. Variables are evaluated based on model performance, which is very situational. The positive words provided the best predictions for \citet{hajek2015}, while the negative words provided best performance for \citet{jo2016}. Even in relation to financial ratios, there is a variation to which ratios best predict bankruptcy. Although the five financial ratios predicted bankruptcy best for Altman, there is no guarantee that they will fit our dataset, and will interact well with our sentiment analysis. As previously mentioned, unfortunately the sector information was not made available to us, meaning that predictions will be somewhat flawed, as we cannot know which sector is which. The four largest available sectors are chosen as our sectors, and each is mapped randomly to the sectors from which we are scraping data.
\\
\newline
As table \ref{Tab:classimb} shows, we have a huge difference in the population size of our two classes.
\begin{table}[H]
\centering
\begin{tabular}{ |c||c|c| }
    \hline
        Year & Bankrupt & Non-Bankrupt \\ \hline
    \hline
        2013 & 9.0\% & 91.0\% \\ \hline
        2014 & 7.3\% & 92.7\% \\ \hline
        2015 & 4.9\% & 95.1\% \\ \hline
        2016 & 3.0\% & 97.0\% \\ \hline
\end{tabular}\par
\caption{Class imbalance rates per year}
\label{Tab:classimb}
\end{table}
This occurrence is commonly referred to as class imbalance, discussed in section \ref{sec:research}. The class imbalance problem is one of the trickiest parts of this experiment. We attempt some different techniques to tackle this problem. SMOTE is used to tackle this issue \citep{chawla2002}. In this technique, synthetic instances are created by the SMOTE algorithm. This is done by working within the feature space, and constructing new instances along line segments joining the minority class $k$ nearest neighbours. SMOTE was used to train our model and generate synthetic examples of bankrupt companies. This experiment was carried out on various algorithms, namely, boosted decision trees, artificial neural networks, and logistic regression. SMOTE was also used in these experiments. Anomaly detection through isolation forests were also attempted to be used, however, returned poor results. All variables mentioned above were used to make our predictions, whilst testing our sentiment variables separately, to try and understand the impact each one has on the model.

ANNs provided some of our better classification results, performing better than other methods out of the box and across all years. SMOTE was performed on the dataset, and this new synthetic data was used to aid the algorithm in recognizing the difference between a bankrupt and non-bankrupt company, and also not showing any bias toward the majority class. Techniques were also evaluated depending on how well they were trained across various years. Boosted decision trees in conjuction with SMOTE were also considered, similar to what had been used by \citet{zikeba2016}. Boosting is described by \citet{schapire2003} as being a method that improves the accuracy of any given machine learning. This method is based on the logic that many simple and slightly inaccurate rules perform can be combined to produce accurate prediction rules. This method provided a variation of results, tending to predict non-bankrupt companies better than ANNs, but also not predicting bankrupt companies as well as ANNs. To evaluate models we consider the accuracy of the model, explained by equation \ref{Eq:acc} below.
\begin{equation} \label{Eq:acc}
    \mathrm{Accuracy}=\frac{TP+TN}{TP+FP+TN+FN}
\end{equation}
However, due to the class imbalance problem, this does not tell the full story. Even a classifier that classifies every company as non-bankrupt will have a strong accuracy. It is far more costly and dangerous to incorrectly score a company going bankrupt that it is safe, rather than vice-versa. For this reason, we evaluate the rate at which our classifier incorrectly classifies bankrupt companies as non-bankrupt. This is referred to as a Type I Error.
\begin{equation} \label{Eq:type2}
    \mathrm{Type\,I\,Error}=\frac{FP}{TN+FP}
\end{equation}
In our experiments we compare the performance of the logistic regression model, with class weights adjusted per year, the Gradient Boosting Classifier and the ANN Classifier. The SMOTE was set to have 4 neighbours, and was used on the Gradient Boosting Classifier and the ANN Classifier. Class weights of logistic regression were adjusted for the various years, set to $9.78$ for 2013 and 2014, $20$ for 2015 and $14$ for 2016. There was no exact science for setting the parameters, and relied somewhat on intuition. The Gradient Boosting Classifier was set to have a learning rate of 0.2 and used 100 stages of boosting. The Neural Network was set to have 2 hidden layers, each with six nodes. All these experiments were carried out using \texttt{Python 3.6} and the \texttt{sci-kit learn} package.
\newline
\begin{table}[H]
\centering
\begin{tabular}{ |c||c|c|c| }
    \hline
        Year & Logistic Regression & Gradient Boosting & ANNs \\ \hline
    \hline
        2013 & 62.7\% & 70.3\% & 52.4\% \\ \hline
        2014 & 76.0\% & 77.3\% & 47.3\% \\ \hline
        2015 & 91.1\% & 84.3\% & 59.4\% \\ \hline
        2016 & 89.6\% & 86.1\% & 64.2\% \\ \hline
\end{tabular}\par
\caption{Accuracy of various machine learning models}
\label{Tab:MLcompacc}
\end{table}

\begin{table}[H]
\centering
\begin{tabular}{ |c||c|c|c| }
    \hline
        Year & Logistic Regression & Gradient Boosting & ANNs \\ \hline
    \hline
        2013 & 45.3\% & 69.8\% & 45.2\% \\ \hline
        2014 & 63.4\% & 68.3\% & 31.7\% \\ \hline
        2015 & 85.3\% & 51.3\% & 33.3\% \\ \hline
        2016 & 82.1\% & 88.2\% & 29.4\% \\ \hline
\end{tabular}\par
\caption{Type I Error of various machine learning models}
\label{Tab:MLcomptype2}
\end{table}
Tables \ref{Tab:MLcompacc} and \ref{Tab:MLcomptype2} illustrate that all three models performed quite differently for our two metrics. The logistic regression model required its weights to be adjusted for each year, since the data was different. In fact, keeping the same class weight from 2013 to 2016 resulted in acceptable results for 2013, but terrible results for 2016. There is a huge gap between the Type I Error obtained by the ANN model and that scored by the others, with the ANN being significantly better at recognizing bankrupt companies. However, the ANN then struggled to identify non-bankrupt companies correctly across all the years, and had a large Type II Error. The Gradient Boosting Classifier improved on the ANNs poor accuracy, but struggled in terms of identifying bankrupt companies correctly. 
\newline
For the purpose of this experiment, we continue working with the ANN. It is certain that with a more detailed amount of tuning and improved feature selection, results would have been better. The SMOTE technique certainly helped the ANN, and should always be considered when carrying out an experiment of this sort. To compare these models, all the sentiment variables were used together. However, we now investigate which variable is having the biggest impact through tables \ref{Tab:neuralacc} and \ref{Tab:neuraltype2}.
\begin{table}[H]
\centering
\begin{tabular}{ |c||c|c|c|c| }
    \hline
        Year & No Sentiment & Negative\% & Positive\% & PositiveToNegative \\ \hline
    \hline
        2013 & 69.6\% & 61.8\% & 53.0\% & 62.8\% \\ \hline
        2014 & 56.7\% & 61.6\% & 68.0\% & 56.6\% \\ \hline
        2015 & 75.1\% & 67.7\% & 59.4\% & 63.8\% \\ \hline
        2016 & 72.3\% & 73.2\% & 73.0\% & 74.9\% \\ \hline
\end{tabular}\par
\caption{Accuracy of Neural Network model}
\label{Tab:neuralacc}
\end{table}

\begin{table}[H]
\centering
\begin{tabular}{ |c||c|c|c|c| }
    \hline
        Year & No Sentiment & Negative\% & Positive\% & PositiveToNegative \\ \hline
    \hline
        2013 & 56.0\% & 59.6\% & 48.0\% & 52.7\% \\ \hline
        2014 & 46.8\% & 52.8\% & 61.0\% & 45.2\% \\ \hline
        2015 & 59.2\% & 65.6\% & 46.7\% & 38.7\% \\ \hline
        2016 & 57.7\% & 63.6\% & 60.9\% & 62.5\% \\ \hline
\end{tabular}\par
\caption{Type I Error of Neural Network model}
\label{Tab:neuraltype2}
\end{table}

\begin{figure}[h!]
\centering
\includegraphics[scale=0.4]{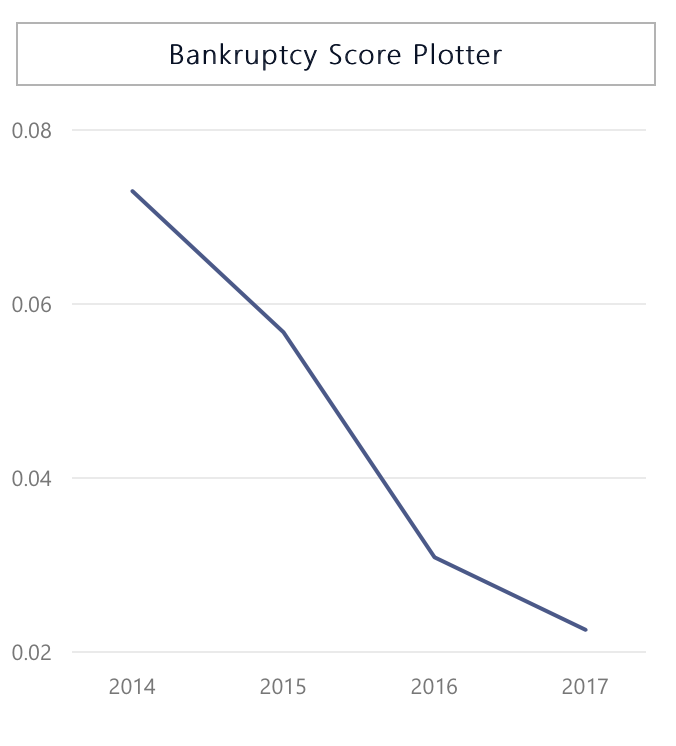}
\caption{The Bankruptcy Score Plotter}
\label{fig:bsp}
\end{figure}

\begin{figure}[h!]
\centering
\includegraphics[scale=0.5]{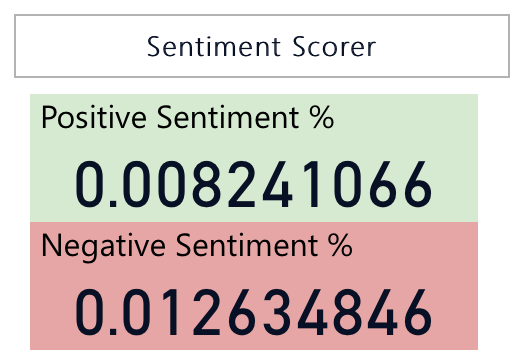}
\caption{The Sentiment Scorer}
\label{fig:ss}
\end{figure}

\begin{figure}[h!]
\centering
\includegraphics[scale=0.5]{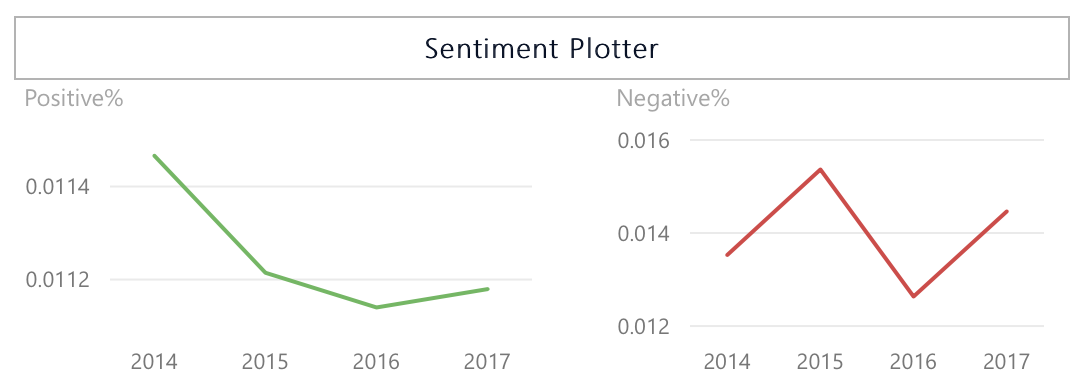}
\caption{The Sentiment Plotter}
\label{fig:sp}
\end{figure}

\begin{figure}[h!]
\centering
\includegraphics[scale=0.5]{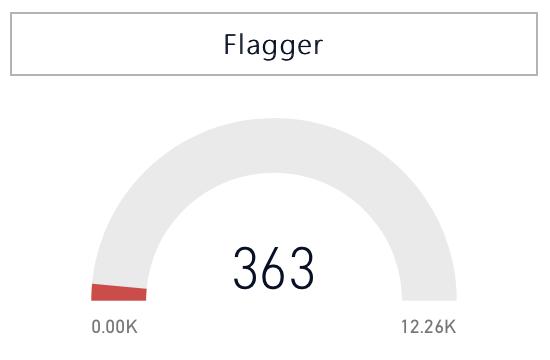}
\caption{The Flagger}
\label{fig:flagger}
\end{figure}

\begin{figure}[h!]
\centering
\includegraphics[scale=0.25]{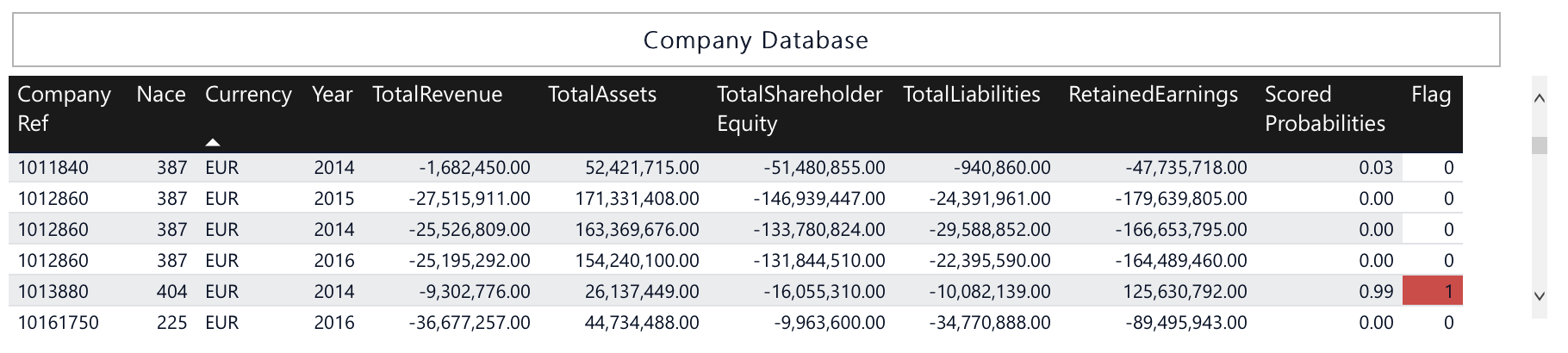}
\caption{The Company Database}
\label{fig:cd}
\end{figure}

The sentimental variable that tended to give us the best predictions was the PositiveToNegative score. However, this was not true for all years, in 2013 and 2015 for instance, the Positive\% score gave the best predictions.  The best results came when using all three variables together, which was done for the experiments illustrated in tables \ref{Tab:MLcompacc} and \ref{Tab:MLcomptype2}. It should also be noted that all three variables returned poor results for 2016. Recall that in table \ref{Tab:classimb} the biggest class imbalance comes in 2016, which may impact the performance of our variables for that year. Not using a sentiment variable and using financial ratios alone resulted in fairly consistent results across years, and in fairly poor results also. Making use of the PositiveToNegative variable resulted in a lower Type I Error across almost all years, except for 2016. Due to data being pseudonymised, this may be a coincidence, however it is interesting to note that predictions were shown to improve with sentiment variables added.
\\
\newline
In general, despite using various techniques and attempting various models, unfortunately a very strong prediction model was not found. Both tables \ref{Tab:neuralacc} and \ref{Tab:neuraltype2}, showed that the variables selected did make a difference, with the Positive\% and PositiveToNegative scores both helping our model predict better than the Negative\% score. As shown in table \ref{Tab:sent}, the PositiveToNegative scores helped encapsulate what is going on within a sector better than each of the other two scores, and it is clear that there is a big difference in the value across sectors and even the trend across sectors. Variable selection clearly needs more investigating, and it is certain that with more time, other financial ratios could be investigated and various combinations of variables will certainly improve the results obtained.
\\
\newline
The final part of this assignment involves proposing a solution to be used by an end user. Below is a description of a dashboard illustrating various statistics related to the selected company or sector. The final dataset containing financial variables and bankruptcy scores is loaded into Microsoft Power BI. This application permits the aggregation of data from different sources to be presented in an interactive manner by means of a visual dashboard. The user is presented with a number of clickable visualizations whereby the information provided can be further filtered, analysed or highlighted. The loaded data contains the company ID numbers; general financial values such as total assets and liabilities, the sentiment analysis scores for each company (based on NACE code), and most importantly, the bankruptcy scores (probabilities) from the trained bankruptcy model. The user is able filter the data according to the company reference ID, the sector, the year or the flagged companies (companies in danger of going bankrupt). All visualizations will change according to the currently selected filters.
\newline
Due to the poor results when using \texttt{Python 3.6}, Azure Machine Learning Studio was attempted to be used to run the machine learning model, however both technologies returned fairly poor results. For demonstration purposes, the training set was included as part of the test set. This will naturally return strong results, as we are testing the set that was trained, however this is being done to show a more realistic view of what the solution should look like. Bankruptcy scores were used instead of classification, so as to allow the user to evaluate the likelihood of bankruptcy according to the score given. Consider a situation where the model assigns a score of 0.51, this would be classified as going bankrupt by the model, however the model is not entirely certain. Similarly a score of 0.99 would be assigned as going bankrupt, however, the model is far more certain. For this reason, the user is left to evaluate based on these scores.
\\
\newline
The dashboard contains five different visualization components, listed below.
\begin{itemize}
    \item \textbf{Bankruptcy Score Plotter}
    \item \textbf{Sentiment Scorer}
    \item \textbf{Sentiment Plotter}
    \item \textbf{Flagger}
    \item \textbf{Company Database}
\end{itemize}
\bigskip
The \textbf{Bankruptcy Score Plotter} (shown in figure \ref{fig:bsp}) illustrates the average bankruptcy score for the entire database per year. With filters, this can be drilled down to a sector or company level. This allows the user to view trends and identify any sectors that may have a greater amount of bankruptcies occurring.
\newline
The \textbf{Sentiment Scorer} (shown in figure \ref{fig:ss}) illustrates the general positive and negative sentiment score related to all companies, or a particular sector, depending on the filters that were set.
\newline
The \textbf{Sentiment Plotter} (shown in figure \ref{fig:sp}) are two graphs showing the average sentiment related to all sectors on a yearly basis, and can be filtered according to the particular sector selected. This will show the user the trend of the general feeling and sentiment around the company based on news articles. 
\newline
The \textbf{Flagger} (shown in figure \ref{fig:flagger}) illustrates the number of companies that have been flagged as being in danger of going bankrupt. This will also give the user an idea of the likelihood of bankruptcy within a sector. A company is being flagged if the scored probability is over 0.98. This is an arbitrary number, that can be changed if a different and more accurate machine learning model is used. 
\newline
The \textbf{Company Database} (shown in figure \ref{fig:cd}) shows the list of companies, general information, the year of assessment, relevant financial figures, flag, and bankruptcy score. This will allow the user to view information related to the company and evaluate the size of the company and what's going on within it.
\\
\newline
\section{Conclusion and Future Work} \label{sec:conc}
In this assignment, our goals were to illustrate the potential of using text mining on newspaper articles in conjunction with financial variables, and to provide a framework for solving this problem. The sentiment analysis scores in table \ref{Tab:sent} illustrate that there is some sort of correlation between the number of positive and negative words in articles related to a sector, and the sector themselves. This is a hugely important outcome, as it strengthens the argument that news articles and other qualitative sources of data can be used to help in predicting bankruptcy. In addition to this, adding the sentiment analysis variables did seem to help in predicting bankruptcy. However, it must be reiterated that these outcomes are somewhat flawed as we do not know if we are comparing the same sectors together. This leads to our first issue within this assignment, which is that without the proper NACE codes available within our dataset, we are taking a leap of faith, and do not know if we are comparing the same sectors. For this reason, we cannot draw conclusions from our results in tables \ref{Tab:neuralacc} and \ref{Tab:neuraltype2}. Training our model to obtain good results proved to be a very difficult task. Different techniques helped, such as SMOTE used in conjunction with neural networks or the gradient boosting algorithm. However, there seems to be a deeper rooted issue to solve this problem related to the variables selected. None of the variables selected could create a situation where we would obtain a strong accuracy and low Type I Error rate.
\\
\newline
To improve this experiment and build upon the work here, there is a huge amount of improvements that can be made. Firstly, having data that is not pseudonymised allows for more concrete conclusions to be drawn from such experiments. In addition, this would allow for a more in depth analysis to be carried out, by looking at how different sentiment analysis variables vary against financial variables or the bankruptcy score itself. Undoubtedly, the selection of more financial ratios should be carried out, and feature selection should be run on all these variables combined to obtain a model that suits the dataset. The financial ratios selected for this assignment clearly did not suit our dataset, hindering the results. In this experiment, we considered just one news source. It would be interesting to consider other news portals and gather an aggregated score from all these portals. Additionally to using more news portals, other sources of information can also improve our predictions. \citet{hajek2015} make use of annual company reports to do this. There are various organizations in Malta, such as the Malta Gaming Authority, the Medicines Authority and the Malta Tourism Authority that release their own annual reports, disclosing facts and figures related to the sector during that fiscal year. Carrying out text mining on such reports and using the Loughran and McDonald dictionary to obtain sentiment scores would be a very interesting exercise. When looking at larger companies, social media sentiment from sites such as Twitter could also be explored.
\\
\newline
The dashboard constructed provides a glimpse to the stage we would like to reach. The ability to browse a database of companies and visualize trends within a sector, both in terms of the sentiment around a company and its likelihood of bankruptcy would be a huge step forward to being able to lower the number of bankruptcies occurring. This would potentially lead to companies being correctly identified as being in danger of going bankrupt, taking the necessary precautions, and potentially saving hundreds, or even thousands of jobs and plenty of money.
\newpage
\bibliographystyle{abbrvnat}
\bibliography{mf_mining_visualizing.bib}

\end{document}